# Hybrid Agentic AI and Multi-Agent Systems in Smart Manufacturing

Mojtaba A. Farahani[a], Md Irfan Khan[a], Thorsten Wuest[a*]

[a]*Mechanical Engineering, University of South Carolina, Columbia, SC, USA*

* Corresponding author. *E-mail address:* twuest@sc.edu

## Abstract

The convergence of Agentic Artificial Intelligence (AI) and Multi-Agent Systems (MAS) enables a new paradigm for intelligent decision-making in Smart Manufacturing Systems (SMS). Traditional MAS architectures emphasize distributed coordination and specialized autonomy, while recent advances in agentic AI driven by Large Language Models (LLMs) introduce higher-order reasoning, planning, and tool orchestration capabilities. This paper presents a hybrid agentic AI and multi-agent framework for a Prescriptive Maintenance (RxM) use case, where LLM-based agents provide strategic orchestration and adaptive reasoning, complemented by rule-based and Small Language Models (SLMs) agents performing efficient, domain-specific tasks on the edge. The proposed framework adopts a layered architecture that consists of perception, preprocessing, analytics, and optimization layers, coordinated through an LLM Planner Agent that manages workflow decisions and context retention. Specialized agents autonomously handle schema discovery, intelligent feature analysis, model selection, and prescriptive optimization, while a human-in-the-loop interface ensures transparency and auditability of generated maintenance recommendations. This hybrid approach enables dynamic model adaptation, transparent decision-making, and cost-aware maintenance scheduling based on data-driven insights. An initial proof-of-concept implementation is validated on two industrial manufacturing datasets. The developed framework is modular and extensible, allowing new agents or domain-specific modules to be integrated seamlessly as system capabilities evolve. The results demonstrate the system's capability to automatically detect schema, adapt preprocessing pipelines, optimize model performance through adaptive intelligence, and generate actionable, prioritized maintenance recommendations. The framework shows promise in achieving improved robustness, scalability, and explainability for RxM in smart manufacturing, bridging the gap between high-level agentic reasoning and low-level autonomous execution.

## 1. Introduction

Manufacturing industries are undergoing a profound transformation in the era of Industry 4.0. Interconnected Cyber–Physical Systems (CPS), fleets of Industrial Internet of Things (IIoT) sensors, advanced AI/Machine Learning (ML) analytics, edge and cloud computing, and digital twin technologies are enabling industries to become more automated, adaptive, and intelligent than ever before. This technological convergence is reshaping manufacturing from reactive and scheduled operations toward data-driven, resilient, and real-time decision-making ecosystems capable of anticipating changes and optimizing performance autonomously [1,2].

The integration of Industry 4.0 technologies unlocks unprecedented capabilities; however, it also introduces new challenges and complexities. Modern manufacturing environments are highly dynamic and data-intensive [3]. Managing data streams from machine tools, processes, and a multitude of sensors in real time presents significant challenges in data fusion, reasoning, and decision making. Traditional rule-based frameworks and centralized predictive analytics pipelines often fail to cope with the scale, variability, and uncertainty of such environments [4,5]. Furthermore, disruptions such as equipment degradation, and production reconfiguration demand adaptive intelligence capable of autonomous diagnosis and response [6].

Predictive and prescriptive maintenance have emerged as key enablers of operational resilience. Predictive Maintenance (PdM) leverages sensor data and AI models to monitor equipment health, detect early signs of degradation, and forecast failures before they occur [7], minimizing unplanned downtime and overall maintenance costs. However, modern manufacturing demands go beyond failure prediction alone, where systems not only anticipate anomalies but also recommend or autonomously execute appropriate corrective actions. Challenges such as scalability, adaptability, interpretability, and latency further highlight the limits of purely centralized PdM approaches [8].

RxM extends predictive capabilities by integrating optimization, reasoning, and decision-support mechanisms to determine the most effective maintenance strategy once a potential issue is detected [9]. Achieving this vision requires distributed intelligence that enables machines, edge devices, and cloud systems to collaborate seamlessly. This need for distributed yet coordinated autonomy forms the foundation for integrating Agentic AI and MAS into next-generation maintenance architectures.

### *1.1 Multi-Agent Systems in Manufacturing*

A promising foundation for distributed intelligence in industrial environments is the MAS paradigm. In MAS, multiple autonomous agents operate semi-independently while coordinating through cooperation, negotiation, or even competition to achieve overarching system objectives [10]. Rooted in the field of Distributed Artificial Intelligence (DAI), MAS emphasizes how complex, system-level behaviors can emerge from the interactions of simpler local entities.

MAS agents are often heterogeneous, and each agent performs *different roles*, have *distinct capabilities* and *pursue varied goals*. Their strength lies in distributed problem solving, coordinated communication, and negotiation, which together allow them to tackle challenges beyond the scope of any single agent. Depending on the application, agents may act as specialists with narrow expertise (e.g., diagnostic, scheduling, or monitoring agents) or as more generalist entities capable of broader tasks.

In manufacturing, MAS has been applied to domains such as production scheduling [11,12], supply chain synchronization, diagnostics, monitoring, and process control [13]. These applications highlight their potential to enhance flexibility, responsiveness, and robustness in cyber-physical production systems. However, the effectiveness of MAS is often constrained by the limited reasoning depth and context awareness of traditional agents [14]. Most frameworks depend on predefined behavioral rules or static task-specific models, requiring extensive manual programming and reducing their adaptability to unforeseen events. As manufacturing environments become increasingly unstructured and data-rich, there is a growing need for agents capable of contextual understanding, strategic planning, and dynamic decision-making beyond fixed procedural logic.

### *1.2 Agentic AI*

Recent advances in foundation language models [15] have given rise to Agentic AI which is defined as autonomous, goal-directed AI systems with *reasoning*, *planning*, *memory*, and *tool-use* capabilities [16]. Unlike classical agents that are constrained by rigid rules or narrow heuristics, agentic AI agents can adapt dynamically to open-ended scenarios, leveraging multimodal inputs, generating strategies under uncertainty, and interacting with external tools or domain-specific knowledge bases. Their ability to plan, reflect, and iterate aligns closely with the evolving demands of complex industrial environments.

Agentic AI systems can parse and interpret multimodal signals (e.g., text, audio, sensor data), reason across uncertain contexts, formulate multi-step intervention plans, and engage in continuous self-improvement [17]. These characteristics make them well-suited for RxM, where agents must not only analyze high-dimensional and noisy sensor streams but also reason about degradation patterns, schedule maintenance tasks proactively, and communicate actionable insights to human operators.

Large Language Model (LLMs) serve as the cognitive core (i.e., the brain) of such agents, excelling at contextual reasoning and complex task orchestration. However, these strengths are accompanied by important trade-offs: LLMs are computationally expensive, prone to latency issues when deployed remotely on the cloud, and raise privacy concerns due to the need to transmit potentially sensitive data to off premise servers [18].

In contrast, Small Language Models (SLMs) offer a lightweight, privacy-preserving complement. Deployed locally on edge devices or factory nodes, SLMs can provide low-latency reasoning, rapid diagnostics, and continuous monitoring without reliance on external infrastructures [19]. Although their reasoning capacity is more limited than LLMs, they are suited for localized tasks such as preliminary fault detection, anomaly flagging, and context-specific decision-making, thereby reducing dependence on centralized resources while maintaining robustness at the edge.

### *1.3 Research Gap and Contribution*

Despite the progress in MAS and agentic AI, their integration remains underexplored. Existing MAS implementations provide distributed coordination but lack the adaptive reasoning of LLM-powered agents. Conversely, current agentic AI systems are often designed as single-agent architectures, limiting scalability and resilience. This gap motivates the development of a proof of concept for hybrid framework that unifies MAS coordination principles with the reasoning-rich autonomy of agentic AI within manufacturing domain applications. By combining MAS autonomy with LLM/SLM-powered reasoning, the proposed framework aims to achieve robustness, adaptability, scalability, and trustworthiness in smart manufacturing environments. Specifically, our contributions are:

- Design of a layered hybrid Agentic AI architecture for predictive maintenance and quality inspection, comprising perception, diagnostic, prediction, and coordination agents.
- Integration of SLMs for local, lightweight, and privacy-preserving reasoning.
- Escalation to LLMs for global, contextual, and complex reasoning when required.
- Implementation of goal-directed, tool-using, and communicative agents, combining distributed autonomy with cooperative intelligence.

## 2. Related Works

Research has demonstrated the benefits of distributed decision-making using MAS in manufacturing systems, but it remains limited by rigid reasoning mechanisms [14,20]. Agentic AI, while initially explored in general-purpose and single-agent prototypes using LLMs, is now experiencing rapid adoption across diverse industrial contexts, including manufacturing and logistics [21]. However, robust industrial applications increasingly rely on MAS and tool-calling architectures to achieve the necessary scalability, specialization, and reliability required for complex, enterprise-level workflows.

In the broader context of smart manufacturing, advanced ML and Deep Learning (DL) methods have also shown promise in RxM. At the same time, these approaches often lack generalization, interpretability, and scalability when applied to dynamic production environments [22]. As a results, hybrid approaches are beginning to emerge, yet research that fully integrates agentic AI with MAS in the manufacturing domain for RxM applications remains in its early stages.

### *2.1 From Predictive to Prescriptive Maintenance*

Although PdM is not yet fully adopted across many factories, it remains an essential goal for SMS. Its adoption is increasing but continues to face practical barriers related to integration costs, data maturity, and organizational readiness. The current trajectory, however, is moving in the direction of autonomous maintenance frameworks in which prediction, diagnosis, and decision-making are integrated within a closed loop. This evolution addresses key limitations of conventional PdM and aligns with the objectives of resilient, scalable, and self-adaptive production systems envisioned in Industry 4.0 and beyond [23].

PdM has emerged as a major research domain in smart manufacturing, driven by the widespread availability of sensor data and the proliferation IIoT in factories (e.g., vibration, temperature, current) [24]. PdM application spans in asset health monitoring, early fault detection, production optimization, quality assurance, spare part management, energy efficiency, and safety enhancement [25].

A variety of statistical and ML approaches have been applied in PdM applications. Classical time-series methods such as ARIMA and exponential smoothing remain in use for baseline forecasting tasks [2]. ML models such as Support Vector Machines (SVM), Linear Regressor (LR), Random Forests (RF), and XGBoost have also been widely used for classification and regression of fault signatures [26]. As the field has advanced, increasingly sophisticated deep learning techniques have been widely adopted. Architectures such as Multi-Layer Perceptron (MLPs) based, Convolutional Neural Networks (CNNs) based, Recurrent Neural Networks (RNNs) based, and autoencoders have been applied to diverse tasks in the PdM domain. For example, signal processing and feature extraction using CNNs, sequential degradation modeling with RNNs, and unsupervised anomaly detection through autoencoders. Benchmark datasets such as NASA's C-MAPSS, the FEMTO bearing dataset, and challenges from the PHM Society continue to serve as standard testbeds for model evaluation [27].

Despite methodological progress, current PdM implementations face several critical challenges that motivate the transition toward prescriptive paradigms. Scalability is a bottleneck: centralizing high-volume sensor streams from many assets creates computational and communication overheads [8]. Adaptability is limited when models trained in one operational regime fail to generalize under distributional shifts or sensor drift [28]. Interpretability is a growing concern, because many high-performance ML/DL models behave as black boxes, undermining operator trust and complicating root-cause analysis [8]. Finally, latency and privacy constraints arise when continuous streaming of raw data to central servers is required. Recent surveys and research trends identify distributed/ federated learning, domain adaptation, and explainable AI (XAI) as active directions addressing these issues [29].

Building on these developments, the emerging paradigm of RxM seeks to close the loop between condition monitoring, prediction, and action. Rather than merely forecasting failures, RxM determines and prioritizes optimal maintenance interventions, integrating diagnostic reasoning, optimization, and decision support within cyber-physical production systems. Recent reviews characterize RxM as an integrative field that leverages ML, reinforcement learning, optimization, and systems integration to move from prediction toward decision-oriented maintenance practice, while noting that conceptual clarity and practical maturity remain incomplete. Large-scale industrial implementation of RxM is constrained by data heterogeneity, the challenge of validating decision models in safety-critical settings, and the need for transparent, verifiable decision logic [30].

### *2.2 Multi-Agent Systems in Manufacturing*

Driven by the need for distributed, adaptive decision-making, research on MAS in manufacturing dates back several decades. Early implementations focused on production scheduling and resource allocation, where agents representing machines, jobs, or work centers interacted to optimize throughput. These systems demonstrated improved scalability, robustness, and resilience to disturbances such as machine breakdowns or supply delays [31].

Subsequent studies expanded MAS applications into maintenance and diagnostics, employing hierarchical or cooperative agent layers for local monitoring and global coordination. Protocols such as the contract-net and auction-based allocation have been used for distributed negotiation and resource sharing [32,33].

Recent studies extend these foundations toward more complex and integrated applications. For instance, a decentralized MAS was proposed to jointly handle scheduling and maintenance planning in flexible job shop environments, enabling agents to negotiate trade-offs between throughput and reliability [34]. Similarly, ontology-based frameworks have been introduced to enrich knowledge representation and improve runtime coordination among manufacturing agents [35]. Parallel to this, the integration of Multi-Agent Reinforcement Learning (MARL) has shown promise in enabling adaptive, data-driven decision-making for manufacturing control systems under uncertainty [36].

More recently, researchers have begun to explore how MAS can be augmented with LLMs and agentic AI to overcome the rigidity of rule-based designs. For example, LLM-enabled agents have been proposed to enhance human–machine collaboration by interpreting natural language instructions, dynamically exploring resource capabilities, and improving system resilience under disruption [37,38]. A recent survey on autonomous and collaborative agentic AI and MAS highlights this convergence as a key enabler of next-generation industrial intelligence [39].

In the context of maintenance, AI-enabled MAS frameworks have also been proposed for predictive and autonomous maintenance, where diagnostic and decision-making agents are combined into a closed-loop architecture that integrates fault detection, classification, diagnosis, and simulation [40].

Collectively, these studies demonstrate that MAS offers a robust foundation for distributed decision-making in manufacturing. However, their reliance on predefined reasoning mechanisms limits adaptability and learning creating an opportunity for integration with agentic AI, which enables dynamic reasoning and continuous improvement.

### *2.3 Agentic AI and Autonomous Agents*

Agentic AI represents a paradigm shift from traditional MAS agents. Agentic AI refers to systems that combine language models such as LLMs with mechanisms for planning, tool use, memory, and self-reflection [41]. Unlike narrowly trained agents, these systems can flexibly interpret multimodal inputs, generate context-aware responses, and autonomously decide when and how to use external tools or APIs.

Recent frameworks such as CrewAI, LangGraph, AutoGen, Semantic Kernel, and MetaGPT have demonstrated the capabilities of agentic AI in domains such as software development, and web-based task automation [42,43]. These frameworks typically follow a plan–act–reflect loop, enabling dynamic strategy adaptation based on feedback. Importantly, they can interact in natural language, making them accessible to non-technical users and supporting intuitive human–agent collaboration.

However, the majority of applications have been non-industrial, focusing on information-centric tasks such as retrieval, report generation, or digital task management in general domains [44,45]. These domains substantially differ from manufacturing, where requirements for safety, reliability, latency, and robustness are significantly stricter. Furthermore, most agentic AI work has emphasized single-agent autonomy rather than distributed collectives, raising questions about scalability in environments such as predictive maintenance where multiple machines and data streams must be managed simultaneously.

In this context, extending agentic AI principles to multi-agent industrial ecosystems represents a promising and unexplored direction in manufacturing domain. Integration agentic reasoning and MAS coordination could enable systems toward fully autonomous maintenance in smart manufacturing.

### *2.4 Hybrid and emerging approaches*

Recent studies have explored hybrid methodologies that integrate MAS with ML techniques. For instance, MAS frameworks have been augmented with reinforcement learning agents that adapt decision rules over time [46], or with anomaly detection models embedded in individual agents [47]. Such hybrids aim to combine the distributed scalability of MAS with the adaptability of ML.

Similarly, digital twin–based predictive maintenance has gained momentum, providing real-time simulations of equipment states for diagnosis and forecasting [48]. However, most digital twin solutions remain centralized rather than distributed, and they lack the conversational, reasoning-driven intelligence characteristic of agentic AI.

To date, the integration of LLM-powered agentic AI into MAS architectures remains unexplored in industrial contexts. This approach has the potential to overcome the rigidity of rule-based MAS while avoiding the bottlenecks of single-agent agentic AI. Each agent can leverage LLM-based reasoning for autonomy and interpretation, while MAS-style coordination ensures scalability and robustness.

In the following section, we describe our proof-of-concept framework bringing the different elements together, with a scenario set in a RxM use case. Section 3 depicts the overall framework, its architecture, key elements including the AI agents, as well as the dataset used for evaluation. Section 4 reports and discusses the results of the framework's evaluation and draws conceptual insights that aim to inspire future research in this domain.

## 3. Methodology

This section describes the hybrid MAS designed for RxM in smart manufacturing. The framework integrates LLMs for strategic orchestration with specialized agents and rule-based tactical components. The description below covers the framework and layered architecture, the end-to-end workflow, key technical components, and implementation details. The source code of the proof-of-concept framework can be found in the paper's GitHub repository at: https://github.com/tamoraji/smart_manufacturing_mas_code.

### *3.1 Framework overview*

We propose a modular MAS in which each agent encapsulates a focused responsibility (data loading, preprocessing, analysis, optimization) and an LLM-based *Orchestrator Agent* provides strategic orchestration. The architecture is hybrid: an LLM handles higher-level reasoning and adaptive strategy selection, while rule-based frameworks and the SLM handle tactical decisions (e.g., model selection, hyperparameter suggestions) for robustness and efficiency. Human-In-The-Loop (HITL) controls are included for review, parameter approval and audit.

### *3.2 Framework architecture*

The framework is structured into five interconnected layers, each managed by specialized intelligent agents with clearly defined roles and interfaces. Each layer communicates with the others through structured inputs and validated outputs, which ensures a seamless, modular, and fault-tolerant workflow.

#### *3.2.1 Intelligence layer (LLM orchestration)*

The LLM *Orchestration Agent* uses an LLM (i.e., gemini-2.5-flash) with reasoning capabilities to reason about workflow state, choose the right tools, and adapt strategies. The orchestrator maintains a workflow context that summarizes goal, available tools, completed steps, the current step index, performance insights, and lessons learned from previous successes or failures. The orchestrator receives a structured prompt that summarizes the *goal*, *toolset*, *step history, performance metrics and any error context*. Using this information, it generates clear and validated instructions for downstream agents. If the orchestrator produces invalid or incomplete outputs, this framework automatically retires up-to *three times* and logs the failed patterns for future improvements. If orchestrator agent is unavailable, then the framework defaults to a rule-based approach. Validation and retry logic ensure the framework remains stable and can recover smoothly from any malformed outputs. Fig. 1 depicts how the *Orchestration Agent* reasons through each stage, selects appropriate tools, and applies rule-based guidance for decision consistency, including adaptive retries and termination checks. These mechanisms ensure that every decision remains transparent, auditable, and aligned with accurate prescriptive goals.

The SLM agents are powered by smaller models that can be run locally (i.e., qwen3:4b via Ollama) to provide quick and cheap tactical suggestions for agents.

### *3.2.2 Perception layer*

The perception layer includes the *Perception Agent*, which is responsible for reading and processing the datasets. This agent computes metadata (incl. shape, column types, missing value counts/percentages, summary statistics) and returns a preview for quick inspection by HITL when transparency is needed or desired. Any data quality issues or anomalies are reported to the LLM *Orchestrator* to inform preprocessing and modelling strategies.

### *3.2.3 Preprocessing layer*

The preprocessing layer is managed by a *Preprocessing Agent*, which automatically builds data preprocessing pipelines based on schema discovery and a tool decider. The agent starts with *schema discovery*, using pattern matching, uniqueness checks, and date and time parsing to label columns as identifiers, timestamps, targets, or features. It then performs feature analysis using statistical methods like correlations, mutual information, and feature importance scores to detect redundancy and low-variance features.

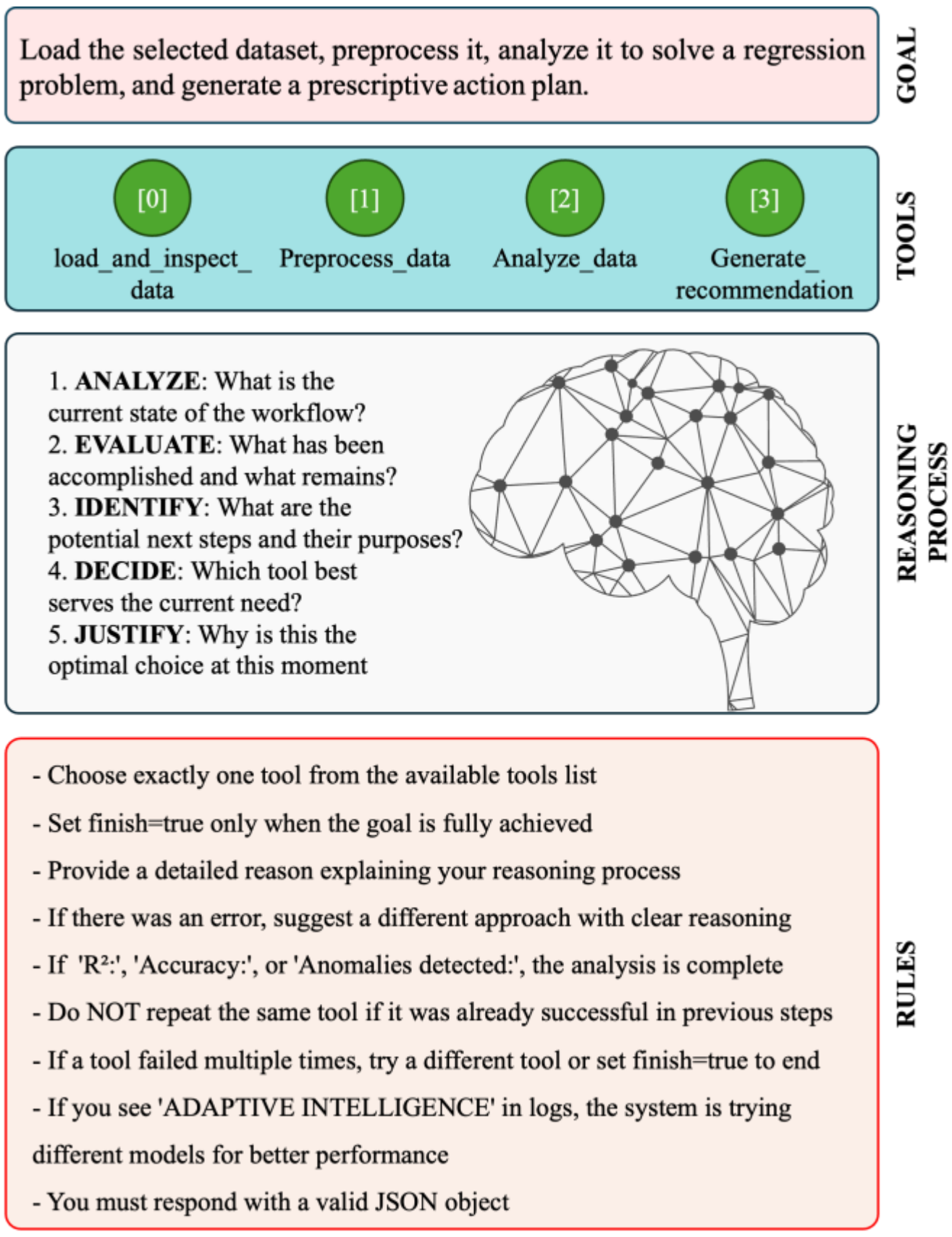

**Fig. 1**. Goal, tools, reasoning process, and rule-based guidance.

Next, the tool decider applies rule-based logic to choose the best preprocessing techniques (imputation, scaling, encoding) based on data traits like missing values, memory size, and category counts. For example, if more than 20% of values in a column are missing, the agent applies a KNNImputer with three neighbours; for moderate missing data (between 10% and 20%), a SimpleImputer using the median is chosen. Datasets with large memory footprints (over 100 MB) trigger the use of a RobustScaler, while smaller datasets use a StandardScaler. Categorical variables with more than 50 unique values may undergo target encoding or be flagged for removal.

Finally, the agent builds a *preprocessing pipeline* where numerical features are imputed and scaled, categorical features are imputed and encoded, identifier columns are passed through unchanged, and the target variable is kept separate to avoid data leakage.

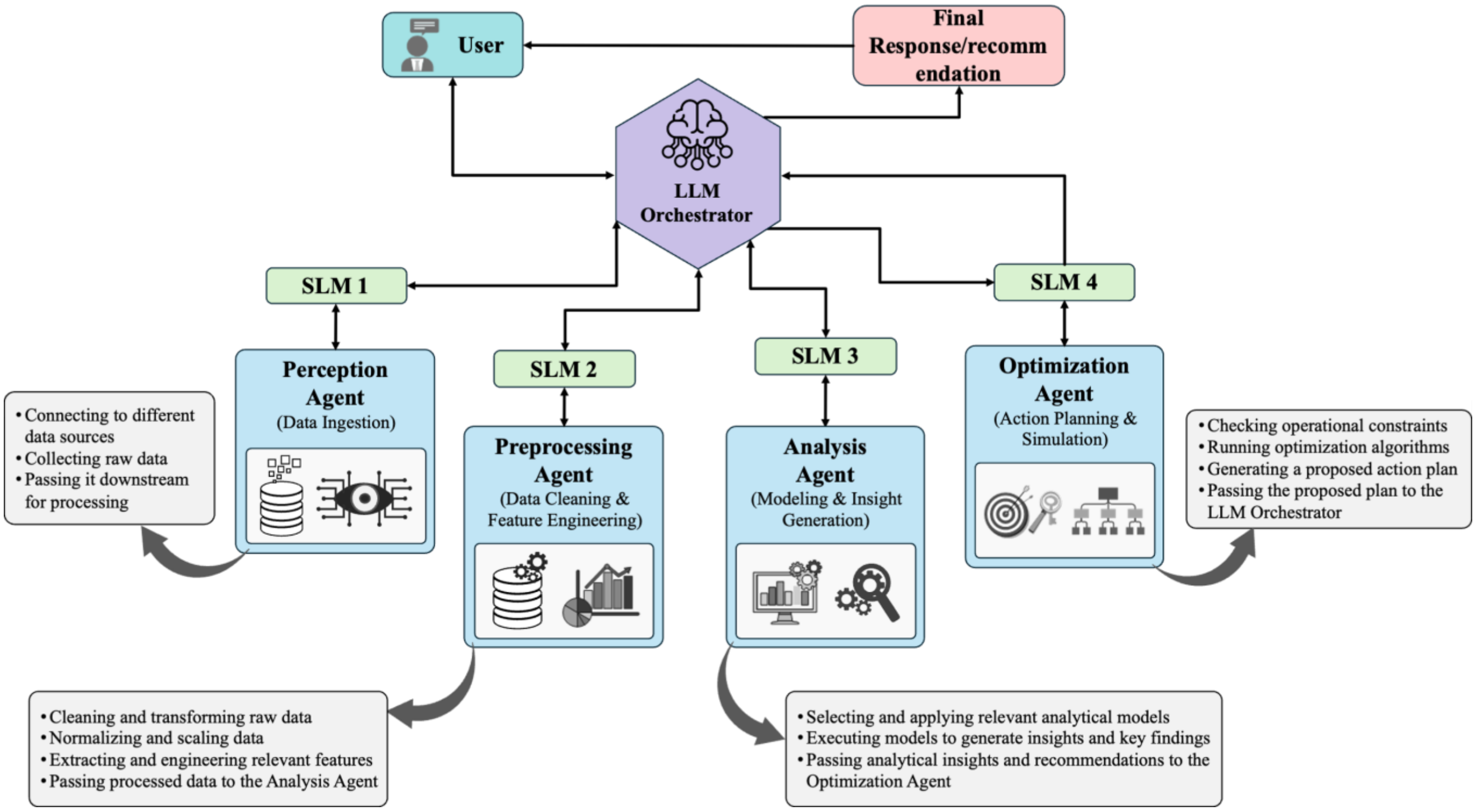

**Fig. 2.** LLM-Driven Multi-Agent Framework Architecture.

### *3.2.4 Analytics layer*

The *Analysis Agent* supports multiple model families, including classification (Random Forest, Logistic Regression, and Support Vector Classifier), regression (Random Forest Regressor, Linear Regression, Ridge, Lasso, and Support Vector Regressor), and anomaly detection (Isolation Forest). During training, the agent performs an 80/20 train–test split and evaluates performance using standard metrics such as accuracy, precision, recall, F1-score, $R^2$, Mean Squared Error (MSE), and anomaly scores.

If performance falls below defined thresholds ($R^2 < 0.1$ or accuracy $< 0.6$), the planner can activate the *Adaptive Intelligence* module, which systematically explores additional model families, records all outcomes, and identifies the best-performing configuration. All model runs, including failed attempts, are logged to ensure transparency, reproducibility, and cumulative learning.

### *3.2.5 Optimization layer*

The optimization layer is powered by the *Optimization Agent,* which translates predictive model outputs and feature importance insights into ranked, actionable recommendations with estimated cost and timeframes. For classification tasks, the agent identifies the top contributing features for each machine using feature importance scores and pre-instance explanations. It then maps predicted priority levels to corresponding actions.

$$Threshold_{high} = mean + 2 \times std \quad (1)$$

$$Threshold_{critical} = mean + 3 \times std \quad (2)$$

$$Priority\ Score = \frac{Predicted\ value - mean}{std} \quad (3)$$

For regression tasks, the agent calculates thresholds based on the training data's mean and standard deviation (Eqn. 1 and 2). Predictions are labelled as warning or critical when they exceed these thresholds, and a priority score is computed (Eqn. 3). For anomaly detection, the agent groups anomalies by machine, computes per-feature z-scores, flags extreme deviations and recommends inspections if multiple features exceed thresholds. Finally, the agent performs a model confidence assessment based on performance metrics. If the confidence score is low, the framework automatically issues warnings to signal reduced reliability

### *3.3 Workflow methodology (end-to-end)*

The framework operates as an end-to-end intelligent workflow that seamlessly integrates data understanding, preprocessing, modeling, and optimization into a coordinated decision pipeline (Fig.2). For prototyping purposes, we utilize an offline dataset as the system's entry point. However, the design is modular, and it allows for future implementations to incorporate a data collection agent that dynamically supplies different data inputs, including, e.g., in-situ machine tool data, to the workflow. The process starts with analyzing the provided dataset to identify column roles, detect data types, compute uniqueness and missingness, and suggest potential target variables and preprocessing strategies. The *Perception Agent* then examines the data's structure, summarizes key statistics, and flags quality issues to inform subsequent preprocessing and modeling decisions.

Using the information from the *Perception Agent*, the *Preprocessing Agent* constructs a dynamic, schema-aware pipeline that performs feature analysis through correlation, mutual information, and feature importance methods. It refines features, removes redundancies, and applies appropriate transformations such as imputation, scaling, and encoding while isolating the target variable to prevent data leakage. The *Analysis Agent* then takes over, selecting suitable models based on dataset characteristics and evaluating them with standard metrics. If performance is unsatisfactory, it automatically explores alternative models and hyperparameters, logging all results for traceability and continuous improvement.

The process concludes with the *Optimization Agent*, which translates model predictions and feature attributions into actionable recommendations ranked by priority. These recommendations include contributing factors, suggested interventions, and estimated cost and time implications. Finally, the framework presents its insights to users for validation, allowing human experts to approve, adjust, or reject the proposed actions. All user interactions are recorded for transparency and auditing, ensuring a reliable and explainable decision-making process.

### *3.4 Technology stack*

The prototype framework was developed in *Python* (version 3.8+) as the primary programming language. Core data processing and ML functionalities were implemented using *pandas*, *NumPy*, *scikit-learn*, and *SciPy*, to support tasks such as data ingestion, feature transformation, statistical analysis, model training, and evaluation.

For intelligent orchestration and reasoning, the g*emini-2.5-flash* LLM was integrated via the Google Gemini API, which serves as the central cognitive engine for high-level workflow planning and decision-making. The LLM Planner Agent utilizes structured prompting and *JSON-based reasoning* traces to determine the sequence of actions, select appropriate analytical tools, and manage inter-agent communication. This integration allows the framework to exhibit agentic reasoning.

To complement cloud-based reasoning with local execution, the framework also employs *Ollama*, an open-source LLM runtime, to host SLMs such as *Qwen3:4B* and *LLaMA3:8B*. These models perform tactical, low-latency decision-making tasks, including preprocessing strategy selection and lightweight analysis steps, enabling the system to operate in environments with limited or no internet connectivity.

Finally, the framework includes an HITL Interface, implemented as a Command Line Interface (CLI) utility with audit logging. The interface enables users to supervise decisions, review reasoning traces, and approve or reject automated actions.

### *3.5 Model selection and hyperparameters*

The framework employs an adaptive approach to model selection, choosing algorithms based on dataset characteristics such as size, dimensionality, and task type (classification, regression, or anomaly detection). Each selected model is trained and fine-tuned using predefined hyperparameters to ensure stable and reproducible performance across different data conditions. The training process uses an 80/20 train-test split, with stratified sampling applied for classification tasks to maintain class balance. The key models and their corresponding hyperparameter configurations are summarized in Table 1.

### *3.6 Metrics and Error handling*

Model performance was evaluated using task-specific metrics. For *classification* tasks, the primary evaluation metric was accuracy, complemented by precision, recall, and F1-scores calculated per class and summarized in a detailed classification report. For *regression* tasks, the primary metric was the coefficient of determination ($R^2$), with MSE serving as a secondary indicator of model performance. In *anomaly detection*, evaluation focused on the number of detected anomalies, corresponding anomaly scores, and per-feature z-scores.

To maintain framework robustness, comprehensive error handling mechanisms were integrated at all stages of the workflow. The framework includes graceful fallback strategies for common failure modes such as JSON parsing errors (handled via automatic retries), model training failures (triggering the next candidate model), preprocessing errors (switching to simpler backup strategies), and data loading issues (returning clear, user-friendly diagnostic messages).

**Table 1.** Model configurations and hyperparameter settings used

| Model | Task | Key Hyperparameters |
|---|---|---|
| Random Forest | Classification | n_estimators = min(100, max(10, n_samples // 10)) |
| Logistic Regression | Classification | max_iter = 1000 |
| SVM | Classification | kernel = 'rbf', C = 1.0 |
| Random Forest Regressor | Regression | n_estimators = 100 |
| Linear Regression | Regression | Default |
| Ridge Regression | Regression | Alpha = 1.0 |
| Lasso Regression | Regression | Alpha = 1.0 |
| Support Vector. Regressor | Regression | kernel = 'rbf', C = 1.0 |
| Isolation Forest | Anomaly Detection | n_estimators = 200, contamination = 'auto' (estimated adaptively if requested) |

### *3.7 Datasets*

Two publicly available manufacturing datasets from Kaggle were used to validate the framework: the *Smart Manufacturing Maintenance Dataset* (SMMD) and the *6G-Enabled Intelligent Manufacturing Resource Dataset (6GMR)*. These datasets represent predictive maintenance and intelligent resource scheduling use cases, allowing evaluation of the system's adaptability across both traditional and next-generation manufacturing contexts.

The SMMD consists of **1,430 records and 10 features**, each representing the operational state of a machine at a specific time. The data integrates real-time sensor readings (temperature, vibration, pressure, and acoustic levels) with maintenance-related variables such as inspection duration, technician availability, downtime cost, and failure probability. The target variable, **Maintenance Priority**, categorizes maintenance urgency into High, Medium, and Low, making it suitable for classification tasks in predictive maintenance research.

The 6GMR contains approximately **100,000 task records and 13 attributes**, capturing the interplay between machine performance and network conditions within a 6G-enabled production environment. It combines equipment-level parameters such as temperature, vibration, and power consumption with network indicators including latency, packet-loss rate, and reliability scores. Additional variables describe product-quality metrics, production speed, and predictive maintenance factors. The target variable, **Efficiency Status**, represents overall process efficiency (high, medium, or low). This dataset supports regression, classification, and anomaly detection experiments aimed at optimizing production performance and diagnosing network-influenced process variations.

Both datasets were selected for their public availability, manageable size thus making them suitable for rapid prototyping, and strong relevance to industrial AI and smart manufacturing research. Together, they enable an evaluation of the hybrid Agentic AI MAS framework across diverse problem types, and decision-making contexts, demonstrating the framework's adaptability to different manufacturing scenarios.

## 4. Result and Discussion

The proposed framework was implemented and validated through a series of proof-of-concept experiments using two representative manufacturing datasets. In total, the complete workflow was executed six times, covering three problem formulations (classification, regression, and anomaly detection) across two datasets. These datasets encompassed diverse operational contexts, including maintenance-priority prediction, process-level performance estimation, and anomaly detection, allowing the framework to be evaluated across multiple analytical problem types.

The primary objective of the experiments was to demonstrate the system's ability to autonomously plan, reason, and execute complete data workflows without explicit task-specific programming and not to optimize prediction accuracy. The results confirm that the hybrid Agentic AI MAS framework can successfully perform end-to-end RxM workflows, generalizes across different datasets and problem formulations, and dynamically selects appropriate tools and models through agentic reasoning and adaptive intelligence.

Furthermore, the framework maintains HITL transparency with minimal technical burden while establishing a scalable and explainable architecture for intelligent manufacturing decision support. Collectively, these outcomes validate the feasibility of integrating agentic reasoning with distributed autonomy to create a unified, modular foundation for next-generation smart manufacturing systems.

### *4.1 End-to-End Workflow Execution*

The framework successfully executed the complete agentic AI workflow across the two datasets. As illustrated in Fig. 3, the LLM Orchestrator coordinated the entire process of data loading, preprocessing, model analysis, and optimization to generate actionable maintenance recommendations with HITL validation.

The workflow consistently followed four agent-led stages: *Perception* (data loading and inspection), *Preprocessing* (feature engineering and quality assessment), *Analysis* (model selection and execution), and *Optimization* (recommendation generation and prioritization). During execution, the LLM dynamically selected agents and tools based on contextual reasoning over workflow state and prior outcomes. Representative structured log entries include:

*"LLM decided: tool='load_and_inspect_data', finish=False, reason='The first step in any data analysis workflow is to load the dataset and inspect its structure and quality.'"*

*"LLM decided: tool='preprocess_data', finish=False, reason='Before analysis, the data must be prepared through scaling, encoding, and intelligent feature selection.'"*

These decisions were generated by the system in real time, not pre-programmed, and the system demonstrated robust error handling and graceful recovery when switching between agents or alternative models. Below, the results for three runs of the workflow are presented.

For the classification task on the SMMD features such as temperature, vibration, pressure, acoustic level, inspection duration, and downtime cost were used to predict maintenance priority (Low, Medium, or High). The *Preprocessing Agent* identified no missing values and standardized the data. The *Analysis Agent* initially selected a RF classifier, and adaptive intelligence confirmed its superior performance over LR and SVC. The model achieved an accuracy of 97.2%, with strong precision and recall across all classes. Feature importance analysis revealed that Downtime Cost and Vibration were the dominant predictors, aligning with domain expectations. Finally, The *Optimization Agent* generated 10 ranked maintenance recommendations. Machines with high downtime cost and vibration intensity were flagged as *Critical*, with suggested immediate inspections and cost/time estimates. For example, it is recommended to immediately dispatch the maintenance team for machine M004 to investigate and restore its performance. Overall, the framework took 132.52s to complete the workflow. Figure 4 illustrates the part of the final prescriptive maintenance output, summarizing prioritized machine actions, cost distribution, and human-in-the-loop approval generated by the agentic AI framework.

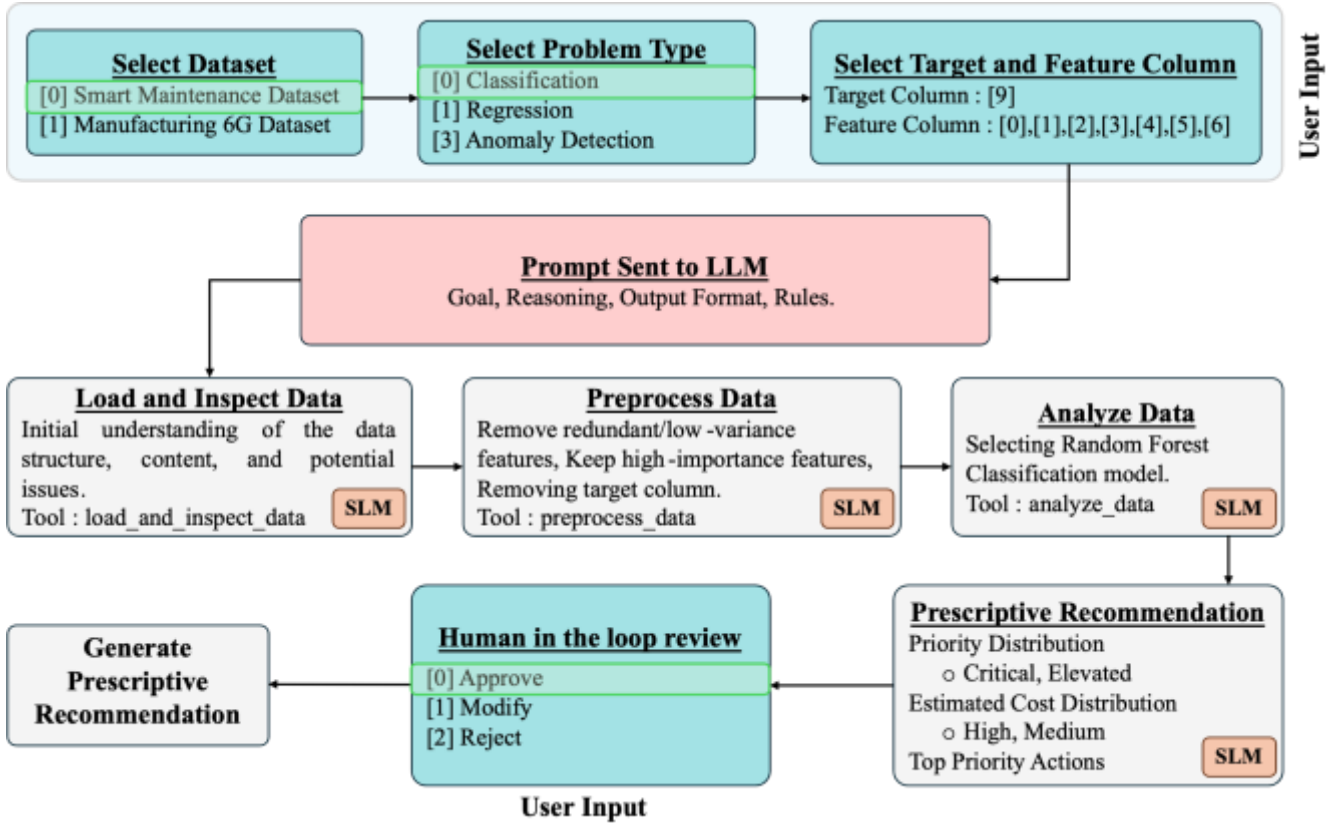

**Figure 3.** Sample agentic AI workflow for LLM-driven prescriptive maintenance using the SMMD.

For the regression task on the SMMD, the *Preprocessing Agent* found no missing values, applied standard scaling and light feature engineering, and identified the top predictors. The *Analysis Agent* selected a Linear Regression model (rule-based choice) which delivered excellent fit with an $R^2$ of 0.9209 and MSE $\approx$ 0.00023. Feature analysis confirmed vibration and temperature as the dominant drivers of the machine failure. The *Optimization Agent* then produced 10 ranked prescriptive maintenance actions (9 Critical, 1 Elevated); top-priority cases include M003, M007 and M004 (predicted failure probabilities $\approx$0.62, 0.61, 0.61 respectively) with recommended immediate inspections and cost/time estimates. The full agentic AI workflow finished end-to-end in 165.26s.

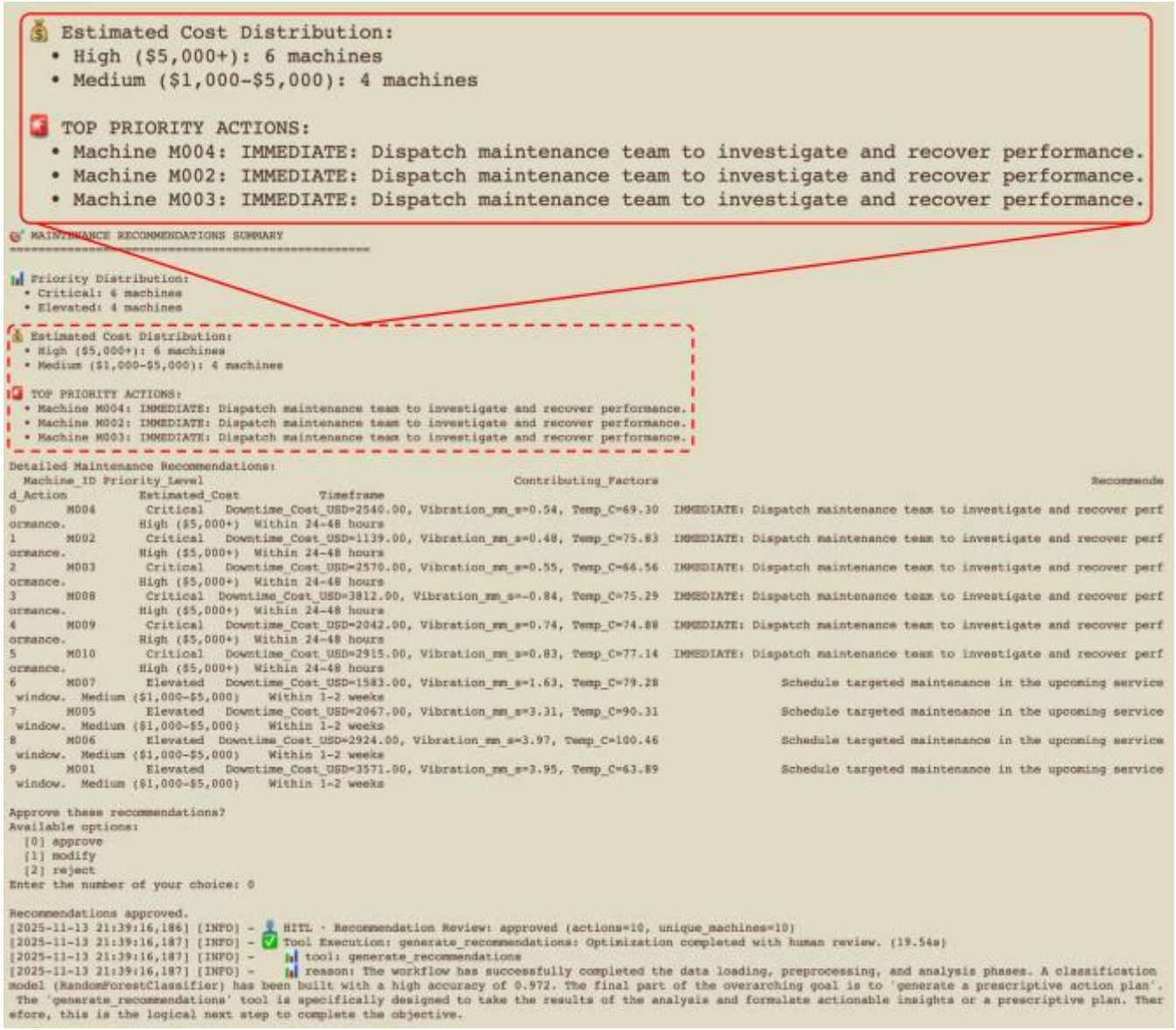

**Fig. 4.** Prescriptive maintenance recommendations generated by the Agentic AI framework for the Smart Maintenance dataset.

For the anomaly-detection task on the 6GMR, the *Preprocessing Agent* verified no missing values, applied standard scaling to the selected input signals, and protected identifier columns. The *Analysis Agent* ran an Isolation Forest (user-tuned: contamination = 0.01, n_estimators = 200) which flagged 1,000 anomalies (1.0% of 100,000 samples). Feature-level z-scores show packet loss and network latency as the primary anomaly signals. The *Optimization Agent* converted detections into 50 advisory prescriptive actions, recommending close monitoring for specific machines with anomaly summaries and suggested observation windows. The agentic workflow completed end-to-end in ~127.4s.

Across all tasks, the framework demonstrated reliable, interpretable, and autonomous orchestration under LLM supervision. Its structured reasoning, rule-based safeguards, and HITL validation collectively ensure traceable decision-making and actionable maintenance recommendation.

### *4.2 Multi-Problem and Multi-Task Adaptability*

One of the core outcomes of this study is the demonstrated adaptability of the framework across distinct manufacturing contexts. Using identical code and configuration, the hybrid system autonomously executed three fundamentally different analytical tasks:

***Classification***: Predicting maintenance priority levels (High / Medium / Low) from sensor and operational data.
***Regression***: Estimating continuous process performance indicators such as error rate or quality metrics.
***Anomaly Detection***: Identifying abnormal operating conditions based on multivariate sensor patterns.

In each scenario, the system autonomously inferred the correct problem type through schema discovery and task reasoning. This ability to handle diverse data structures and problem formulations without manual reconfiguration demonstrates task-agnostic intelligence and generalizability which is the key differentiator from conventional MAS or static ML pipelines.

### *4.3 Agentic Reasoning and Adaptive Intelligence*

A defining feature of the framework is its use of agentic reasoning to coordinate inter-agent decisions. The LLM Planner Agent used *chain-of-thought style deliberation* to select each tool, justify its choice, and anticipate next steps. Each reasoning cycle produced a transparent, machine and human readable explanation, which enables interpretable orchestration of the MAS workflow.

In several runs, the system invoked the *adaptive intelligence* mechanism. A meta-reasoning process where the analysis agent automatically retried multiple models when initial performance metrics were suboptimal. For instance, in regression mode, the agent sequentially evaluated models such as Linear Regression, Ridge, and RF Regressor, ultimately selecting the most stable configuration based on contextual feedback. This behavior exemplifies the framework's ability to perform reflective model adaptation without explicit retraining instructions, fulfilling the principles of self-evaluating agentic systems.

Moreover, the combination of LLM for high-level orchestration and SLMs for local execution provided a balanced trade-off between reasoning depth and computational efficiency. SLMs operating on local nodes handled lightweight preprocessing and feature-analysis tasks with minimal latency, while the LLM Planner coordinated complex reasoning sequences that required contextual understanding or workflow adaptation. In additional, using local SLMs may also present potential opportunities for handling sensitive analytics tasks locally to align with cybersecurity measures.

### *4.4 Human-in-the-Loop Transparency*

Human oversight in the framework serves as a supervisory mechanism rather than a technical requirement. In an environment like manufacturing, retaining a level of human subject matter expert control is highly desirable, especially during the early stages of implementing agentic systems. The HITL Interface logs every agent decision, reasoning trace, and recommendation in real time, producing an auditable record that can be reviewed or overridden by a domain expert or operator. Importantly, the interface abstracts away ML complexity, allowing users without ML or programming expertise to oversee the system through structured summaries. An example of summarized report generated by the framework is shown in Fig 5.

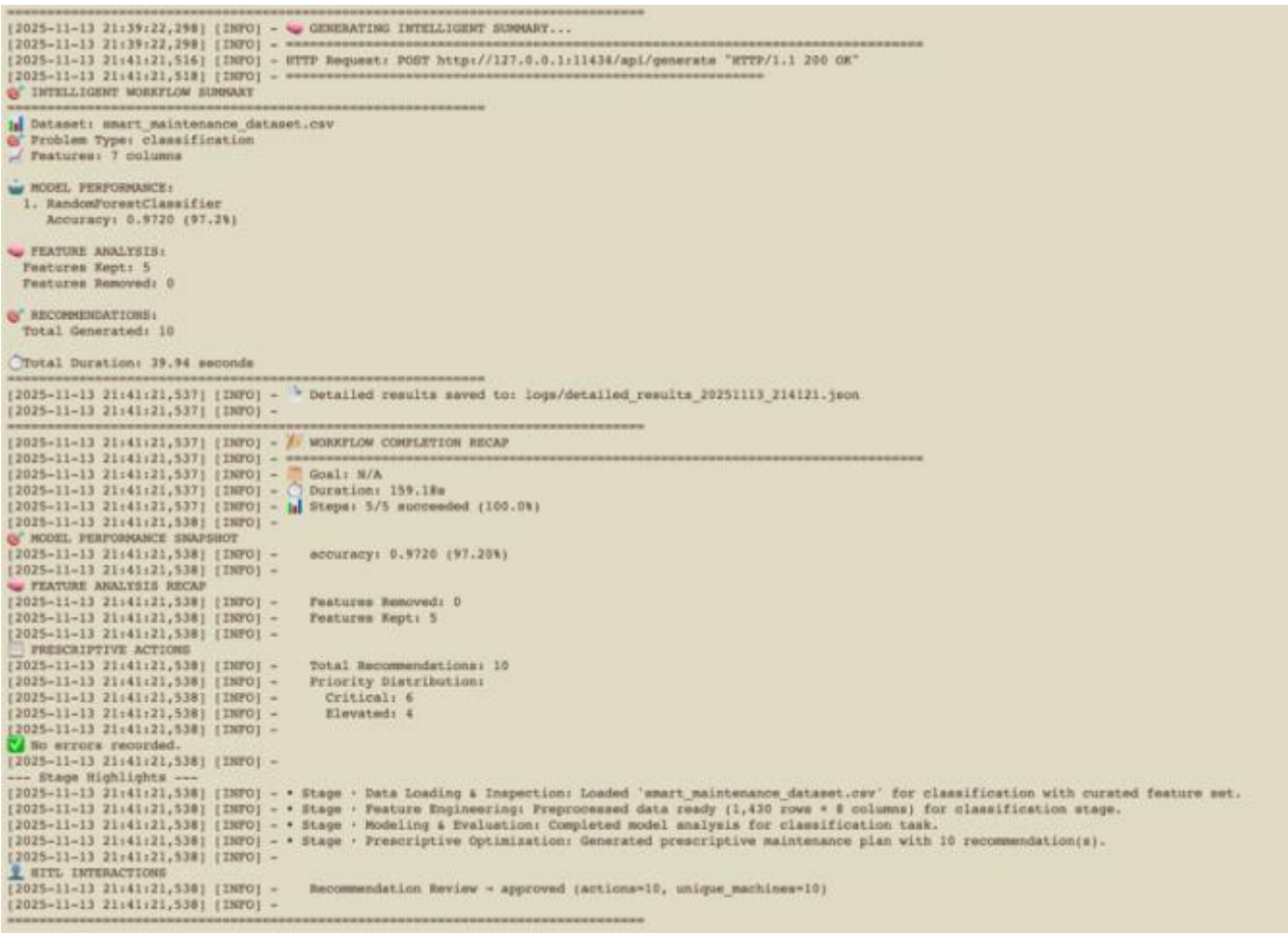

**Fig. 5.** Intelligent summary report generated by the Agentic AI framework for the Smart Maintenance dataset.

### *4.5 Conceptual Insights and Discussion*

The experimental validation confirms that the hybrid Agentic AI MAS framework effectively bridges the gap between centralized reasoning and distributed execution. Conceptually, several key insights emerged from this work.

Integrating agentic AI reasoning into a MAS architecture enables both contextual intelligence and modular autonomy. Within this configuration, the LLM functions as an orchestrator responsible for high-level decision orchestration, while the individual agents specialize in well-defined operational domains such as data preprocessing, analysis, or optimization. This hybrid synergy allows the system to combine broad contextual reasoning with the efficiency and specialization of distributed agents.

The framework also demonstrates notable *scalability* and *extensibility*. Its modular design supports the integration of new agents without requiring re-engineering of the core planner logic. This flexibility ensures that the architecture can evolve alongside emerging industrial applications and computational tools.

*Explainability* is another key outcome of the system design. Each agentic decision is recorded alongside its reasoning trace, transforming what would otherwise be opaque LLM-driven choices into transparent, auditable processes. These structured justifications strengthen trust in automated decision-making and facilitate human understanding of how conclusions are reached.

Equally significant is the framework's capacity for task *generalization*. It demonstrated robust adaptability across diverse data types, learning objectives, and manufacturing contexts without the need for retraining or task-specific scripting. This capability highlights the system's potential to operate as a universal analytical framework for smart manufacturing scenarios.

Finally, the framework maintains a human-centric orientation through its transparent supervisory interface, which ensures that the ultimate decision authority and full control remain with human operators. By allowing shop-floor personnel with limited data and ML expertise to oversee and validate automated actions, the system reinforces accountability, trust, and confidence in hybrid human–AI collaboration.

## 5. Conclusion, Limitation, and Future Work

This paper presented a hybrid Agentic AI and MAS framework designed to advance intelligent decision-making within smart manufacturing environments. By integrating the reasoning and planning capabilities of LLMs with the modular autonomy of MAS, the proposed architecture enables distributed, goal-directed intelligence that can plan, adapt, and act across heterogeneous manufacturing scenarios. The proof-of-concept implementation demonstrated that the framework can autonomously perform complete analytical workflows across multiple datasets and problem types, including classification, regression, and anomaly detection.

The experiments confirmed that the framework can reason about problem context, select appropriate analytical tools, and coordinate specialized agents without explicit programming or manual intervention. This capability represents a fundamental shift from traditional rule-based MAS designs toward systems that exhibit agentic reasoning where agents can plan, justify, and adapt their decisions dynamically. The results also highlight the importance of human-centered transparency that every decision is logged, explained, and reviewable through a HITL interface, ensuring interpretability, transparency, control, and trust even as autonomy increases.

Beyond technical feasibility, this work illustrates the conceptual potential of merging high-level cognitive reasoning with distributed operational autonomy in industrial AI. The hybrid Agentic AI MAS approach provides a scalable and explainable foundation for future intelligent manufacturing systems that can operate with minimal supervision, adapt to new data sources and tasks, and integrate seamlessly with edge, cloud, and digital-twin infrastructures.

*Future work* will focus on advancing the framework's intelligence, autonomy, and integration with real manufacturing environments. The perception layer will be expanded to include intelligent perception agents capable of understanding the nature and structure of incoming data, automatically identifying data types, and suggesting potential analytical problems and corresponding solution strategies. This enhancement will also enable the system to ingest and interpret streaming data directly from industrial equipment through real-time communication protocols such as MQTT and OPC UA, facilitating seamless integration with existing factory infrastructures. Further developments will introduce agent-to-agent communication mechanisms, allowing agents to share context, coordinate reasoning, and make collaborative decisions to improve overall system intelligence. The Analysis Agent will be augmented with state-of-the-art machine learning and optimization algorithms to enhance analytical performance and predictive accuracy across diverse tasks. In addition, the framework will be extended to support multimodal input data for richer contextual understanding. Finally, beyond generating prescriptive recommendations, the system will evolve to autonomously perform operational actions under user supervision and approval, moving closer to fully autonomous, human-aligned manufacturing intelligence.

While the proof-of-concept implementation confirms the framework's feasibility, several *limitations* should be acknowledged. Current system latency during LLM reasoning steps may restrict real-time responsiveness, particularly in streaming data environments. Future optimization of local LLM deployment and SLM orchestration will help mitigate dependency on cloud-based inference services. Moreover, the framework was validated only on two offline datasets, which do not fully capture the dynamic, high-volume nature of industrial data streams. Real-time scalability, latency performance, and robustness under continuous IIoT data remain open challenges. The system's reliance on specific model configurations may also affect performance and cost across different language models or deployment settings. Additionally, while HITL mechanisms enhance transparency, LLM-driven reasoning may still produce inconsistent outputs in unseen scenarios, underscoring the need for stronger safety verification. Finally, privacy and data governance considerations for cloud-based inference were not deeply explored.

Overall, this study demonstrates that the convergence of Agentic AI and MAS principles can yield intelligent, flexible, and transparent systems capable of bridging the gap between reasoning and execution in manufacturing. The proposed framework represents an essential step toward realizing autonomous, human-aligned decision-making ecosystems that embody the vision of Industry 4.0 and beyond.

## Acknowledgements

This material is based upon work supported by the National Science Foundation (NSF) under Grant No. 2119654 & 2519581. Any opinions, findings, and conclusions or recommendations expressed in this material are those of the author(s) and do not necessarily reflect the views of NSF.